\documentclass{article}
\usepackage{graphicx}
\usepackage{amsmath}
\usepackage{amsfonts}
\usepackage{amssymb}

\begin{document}

\title{Towards artificial muscles}
\author{P.\ G.\ de Gennes\\Coll\`{e}ge de France, 11 place M.\ Berthelot\\75231 Paris Cedex, France\\e.mail: pierre-gilles.degennes@espci.fr}
\maketitle
\begin{abstract}
Solid state actuators (piezoelectric, ferroelectric, ...) deform under an
external field, and \ have\ \ found many applications. They respond fast, but
their mechanical deformations are very small.\ There is a need for soft
actuators, giving larger responses, but necessarily less fast.\ This paper
describes the early attempts by Katchalsky and others, and the problems which
showed up -related to time constants, and, most importantly, to
\ fatigue.\ Two current attempts are reviewed.

\medskip

\textit{Key words}: actuators,\quad\ artificial muscles,\quad\ nematic
polymers,\quad\ electro osmosis.

\medskip

\textit{PACS numbers}: 64.70 Md, \quad65.70 +y,\quad\ 66.30 Qa.
\end{abstract}

\section{Introduction}

\begin{figure}
[h]
\begin{center}
\includegraphics[
height=1.67in,
width=2.9776in
]%
{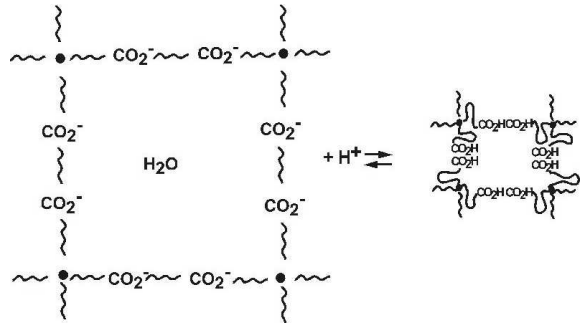}%
\caption{The simplest mechanochemical device: going to acid \emph{pH} causes
contraction.}%
\label{fig1}%
\end{center}
\end{figure}

As soon as Kuhn understood the flexibility of polymer chains, and the origin
of rubber elasticity, his student A.\ Katchalsky thought about the possibility
of transforming chemical energy into mechanical energy, using gels swollen by
water.\ His first idea is explained on Fig.\ref{fig1}. Starting from chains
which carry acid groups $(CO_{2}H)$ and adding $OH^{-}$ ions, one obtains a
charged network $(CO_{2}^{-})$ where the chains stretch by electrostatic
repulsions.\ If one then adds $H^{+}$, the system returns to neutral, and the
gel contracts.%

This system, however, does not allow for many cycles. Adding $OH^{-}$ really
means adding soda $(NaOH)$ and adding $H^{+}$ means adding hydrochloric acid
$(HCl)$.\ At each cycle, one thus adds one mole of $NaCl$, and this ionic
solute screens\ out the electrostatic interactions: the system dies out fast.

Katchalsky solved this problem by an intelligent trick: he used ion exchange
$(Na^{+}$ against $Ba^{++})$ where $Ba^{++}$ binds\ two $(CO_{2}^{-})$ groups
and contracts the gel.

His group produced active fibers of this type\cite{katchalsky55} $.$ However,
the process did not gave rise to useful applications, for a number of reasons:

1) Time constants: what is implied here is diffusion of ions\ from a bath to a
fiber, and diffusion is always very slow. And even if the ions were injected
locally (by highly divided electrodes, or, by conducting polymers) the
diffusion of water remains necessary to swell or deswell.

2) Fatigue: if we swell a gel by water, the swelling process starts at the
outer surface, and creates huge mechanical tensions in a thin region: buckling
instabilities occur$\cite{tanaka87}$ and fractures show up.\ There is little
hope for an artificial muscle which breaks locally at each cycle.

In the following section, we present two attempts where these difficulties are
taken into account.\ It may well be that none of them gives a durable answer,
but the trends are interesting.

\section{A semi fast nematic muscle}

\begin{figure}
[h]
\begin{center}
\includegraphics[
height=1.324in,
width=2.9326in
]%
{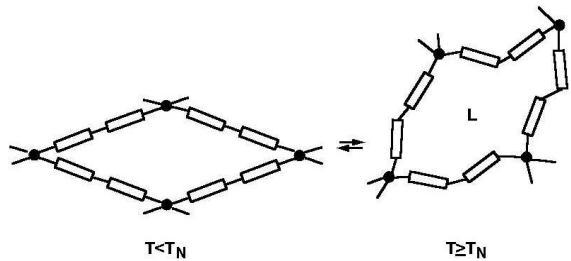}%
\caption{A nematic network: below the ordering point, the network
(\textit{T$<$T}$_{N})$ is elongated.\ Above (\textit{T$<$T}$_{N})$ (in the
isotropic phase) it contracts.}%
\label{fig2}%
\end{center}
\end{figure}

The starting point here is a nematic network schematized on Fig.\ref{fig2}. At
low temperatures, the system is elongated.\ At higher temperatures, above the
nematic isotropic transition point $T_{N1}$, the network contracts.\ Networks
of this type have been synthetised by a very intelligent technique
\cite{finkelman81}.%

Of course, thermal effects have their difficulties: the diffusion of heat is
faster than the diffusion of solvents, but is still slow.\ This led us to
think about a \textit{semi-fast }system\cite{degennes97}$:$ here we heat up
rapidly the system by a light pulse (having some adsorbing dyes inside) and
induce\ the nematic isotropic transition.\ The contraction time of rubber in a
sling -is related to the velocity of shear waves in the rubber, and is
typically of order 1 millisecond.\ When we want to close the cycle, we have to
cool down the sample (by a few degrees) and this takes a long time
(seconds).\ But this semi fast actuator might be of some use.

\begin{figure}
[h]
\begin{center}
\includegraphics[
height=1.7417in,
width=2.7233in
]%
{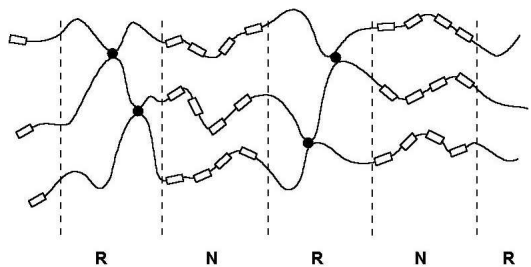}%
\caption{A lamellar phase of block copolymers with rubber portions \textit{R}
and nematic portions \textit{N}.\ The $N$ portions are temperature sensitive
as in fig. 2}%
\label{fig3}%
\end{center}
\end{figure}

To avoid fatigue, we conceived a system based on block copolymers
(Fig.\ref{fig3}) which is hopefully well protected. The fabrication of these
sophisticated copolymers requires artistic chemistry, and is under
way\cite{auroy}$.$%

\section{Nafions}

\begin{figure}
[h]
\begin{center}
\includegraphics[
height=1.7296in,
width=1.8031in
]%
{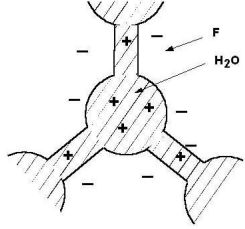}%
\caption{A simplified view of wet nafion membranes. The matrix is a
fluorinated hydrocarbon network.\ At the surface, fixed sulfonated groups bing
a negative charge.\ In the water pockets, the counterions (+) are mobile.}%
\label{fig4}%
\end{center}
\end{figure}

A completely different approach has been used by M.\ Shahinpoor and coworkers
\cite{shahinpoor98}$.$ Here, the basic material is commercially available,
cheap, and robust.\ It is a ''nafion'': a fluoropolymer containing some fixed
$SO_{3}^{-}$ groups plus $Na^{+}$ counterions and small water pockets which
are inter connected (Fig.\ref{fig4}).\ Shahinpoor was able to set up
electrodes, with large contact areas, on both sides of a thin nafion sheet, by
formation of platinum nanoparticles.%

\begin{figure}
[h]
\begin{center}
\includegraphics[
height=1.7002in,
width=2.1586in
]%
{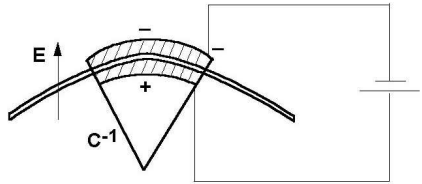}%
\caption{A typical mechanoelectric experiment with a wet nafion membrane}%
\label{fig5}%
\end{center}
\end{figure}

When such a sheet is put under a mild voltage ($\lesssim$1 volt) it deforms as
shown on Fig\ref{fig5}.\ The basic process appears to be simple.\ When a
$Na^{+}$ ion moves, it drags some water with it towards the cathode: thus the
cathode swells, while the anode contracts.%

There is also an inverse effect: if a bending torque\ is applied to the sheet,
a voltage difference shows up.\ 

The features can be described rather simply (like all electro osmosis effects)
in terms of two coupled fluxes (normal to the plate): the electric current $J$
and the water flux $Q\cite{degennestobe}.$ The corresponding forces are the
electric field $E$ and the pressure gradient $\nabla p:$%

\begin{equation}
J=\nabla E-L_{12}\nabla p
\end{equation}%

\begin{equation}
Q=L_{21}E-K\nabla p
\end{equation}

Here, $\sigma$ is the conductivity, $K$ the Darcy permeability and
$L_{12}=L_{21}=L$ is a coupling coefficient.

1) The direct effect corresponds to $Q=0$ and $\nabla p=L/KE.\;$The pressure
gradient induces a curvature $C$: in the absence of any external toque, they
are related by:%

\begin{equation}
\nabla p=kYC
\end{equation}

where $Y$ is the Young modulus, and $k$ is a numerical factor involving the
Poisson ratio $\sigma_{p}$:%

\begin{equation}
k^{-1}=(1+\sigma_{p})(1-2\sigma_{p})
\end{equation}

2) For the inverse effect, we have $J=0,$ and $E=L_{12}\nabla p$ is
proportional to the curvature.

One can construct order of magnitude estimates for the various
coefficients.\ For the very small pores of interest here, the standard
(Smoluchow ski) description of electro osmosis is not very adequate.\ The
following is an alternate view point:

a) The conductance is:%

\begin{equation}
\sigma=\frac{ne^{2}}{\zeta}%
\end{equation}

where $n$ is the number of sodium ions per unit volume and $\zeta$ a friction coefficient:

b) The Darcy permeability $K$ is of order:%

\begin{equation}
K\sim\phi\frac{d^{2}}{\eta}%
\end{equation}

where $\phi$ is the water volume fraction, $d$ the size of the water pores,
and $\eta$ the viscosity.

c) Finally, the coupling coefficient $L$ can be estimated from a situation
where $\nabla p=0$, assuming that each $Na^{+}$ ion drags a volume $w$ of
water. Then:%

\begin{equation}
Q=LE=\text{v}w
\end{equation}

where v=$eE/\xi$ is the drift velocity of the ions. It seems that, in this
way, we can arrive at a reasonable picture of the whole effect.

We are currently studying the time response, ie the frequency dependence of
all these processes. But some general features emerge:

1) The system is robust, and can be cycled many times.

2) The frequency response is typically in the range of 30 cycles -ie slow, but
still of possible interest for some biomedical applications.

On the whole, it is clear that soft actuators are still in their infancy; but
it is also clear that they will soon become important.

\bigskip
\end{document}